\documentclass[aps,pra,preprint,groupedaddress]{revtex4}
\usepackage{amssymb}
\usepackage{epsfig}
\usepackage{color}
\usepackage{amsmath}

\def\beq{\begin{equation}}
\def\eeq{\end{equation}}
\def\bea{\begin{eqnarray}}
\def\eea{\end{eqnarray}}
\def\ba{\begin{array}}
\def\ea{\end{array}}

\def\ketz{|0 \rangle}
\def\keto{|1 \rangle}
\def\sour {|s \rangle}
\def\asour {\langle s|}
\def\targe{|t \rangle}
\def\atarge {\langle t|}

\begin{document}

\title{Simulated Quantum Computation of Global Minima}
\author{Jing Zhu, Zhen Huang and Sabre Kais\footnote{kais@purdue.edu}}
\affiliation{ Department of Chemistry and Birck Nanotechnology Center,
Purdue University, West Lafayette IN 47907}

\begin{abstract}
Finding the optimal solution to a complex optimization problem
is of great importance in practically all fields of science,
technology, technical design and econometrics.
We demonstrate  that a modified Grover's
quantum algorithm can be
applied to real problems of finding a global minimum
using modest numbers of quantum
bits. Calculations of the global minimum of simple test functions
and Lennard-Jones clusters have been carried out on a quantum
computer simulator using
a modified Grover's algorithm. The number of function evaluations $N$ reduced
from $O(N)$ in classical simulation to $O(\sqrt{N})$ in quantum simulation.
We also show how the Grover's quantum algorithm
can be combined with the classical Pivot method for global optimization
to treat larger systems.

\end{abstract}

\maketitle

\clearpage Rational drug design, molecular modeling, quantum
mechanical calculations and mathematical biological calculations are
but a few examples of fields that rely heavily upon the location of
a global minimum in a multiple-minima
problem\cite{sc-kirkpatrick,sc-barhen,sc-cvijovic,sc-wales,NIGRA,KAIS}. Several
global optimization methods have been developed over the past
decades. However, the large computational cost of finding the global
minimum for large number of variables  limited the applications of
such algorithms\cite{Goldberg,Frantz,Kais1-Opt,Kais2-Opt}. Quantum
algorithms on the other hand known to speed up the computation
compared to classical ones\cite{Lioyd1,Lioyd2,Farhi,Cleve}. For
example, the calculation time for the energy of atoms and molecules
scales exponentially with system size on a classical computer but
polynomially using quantum algorithms\cite{Alan,Wang}.

Quantum computation is generally regarded as
being more powerful than classical computation. The evidence for
this viewpoint begins with Feynman's pioneering observation
\cite{feynman} that the simulation of a general quantum evolution
on a classical computer appears to require an exponential overhead
in computational resources compared to the physical resources
needed for a direct physical implementation of the quantum process
itself. Subsequent work by Deutsch \cite{deutsch}, Bernstein and
Vazirani \cite{bv}, Simon \cite{simon}, Grover \cite{grover}, Shor
 and others\cite{shor,shor2} showed how quantum evolution can be
harnessed to carry out some useful computational tasks more
rapidly than by any known classical means. For some computational
tasks (such as factoring) quantum physics appears to provide an
exponential benefit, but for other tasks (such as NP complete
problems \cite{gj}) the quantum benefits appear to be inherently
more restricted, giving at most a polynomial speedup
\cite{bbbv,jozsa1,jozsa2,ekertjozsa,Levine,Munich}.

Grover's quantum algorithm can find an object in a unsorted database
containing $N$ objects in $O(\sqrt{N})$ quantum mechanical steps
instead of $O(N)$ steps\cite{Grover97,Grover05}. The steps of
Grover's Algorithm can be shown as following: firstly, the
Walsh-Hadamard transformation was performed and the system was
initialized to the superposition. Before going any further, let's
introduce some fundamental information of quantum simulation
\cite{quantumBook}. Quantum computation and quantum information are
built upon the concept the \emph{quantum bit}, or \emph{qubit}
briefly. It is quite similar to bit in classical computation. Two
possible states for a qubit are the states $\ketz$ and $\keto$,
which is similar to state 0 and 1 for a classical bit. The
difference between bits and qubits is that a qubit can be in a state
other than $\ketz$ or $\keto$. It is also possible to form linear
combination of states, called \emph{superpositions}: \beq
 |\psi \rangle \,=\, \alpha \ketz \,+\, \beta \keto
 \eeq where $\alpha$ and $\beta$ are complex numbers and
 $|\alpha|^2+|\beta|^2=1$. Furthermore, one famous qubit gate is \emph{Hadamard gate}
 , it is defined as: \beq H \equiv \frac{1}{\sqrt{2}} \left(
\begin{array}{cc}
1 & 1\\
1 & -1
\end{array}
\right) \eeq This gate is also described as being like a
'square-root of not' gate, in that it turns a $\ketz$ into $(\ketz +
\keto)/\sqrt{2}$ (first column of H), 'halfway' between $\ketz$ and
$\keto$; and turns $\keto$ into $(\ketz - \keto)/\sqrt{2}$ (second
column of H), which is also 'halfway between $\ketz$ and $\keto$. In
our simulation, we performed the Hadamard transform on n qubits
initially in the all $\ketz$ state and get an equal superposition of
all computational basis states. Mathematically it could be expressed
in the following formula: \beq
  \sour \,=\,
  \overset{n}{\underset{i=1}{\bigotimes}}
 H \ketz  \eeq

 where $\sour$ means the initial
 superposition and n means the number of qubits. Secondly, we
 generated two different operators called $P_{s}$ and $P_{t}$. Their
 mathematical formations are $P_{t}=I-2\targe {\langle t|}$ and $P_{s}=2\sour {\langle
 s|}-I$, where $\sour$ is the initial superposition and $\targe$ corresponds to the entry
 matching the search criterion. In practice, there are no direct
 universal quantum algorithms currently to obtain $\targe$ besides
 `black box function' or `oracle' \cite{quantumBook, quantumBook1}. In order to get $\targe$ here, we first did classical
 comparison based on search criterion and then translated the result into the quantum state.
 The detailed procedure is provided in the Appendix.
 The physical meaning of
 $P_{t}$ is performing the selective phase inversion, and that of
 $P_{s}$ is performing the inversion about average operation, which
 increases the amplitude of the state which was inverted in the
 previous step. Finally, we applied the Grover operator, $G=P_{s}P_{t}$, $O(N^{1/2})$
  times to the superposition state. After performing the measurement to
  the obtained vector, the one with the max probability is the marked
  state \cite{Grover97, Grover05}. Grovers' search algorithm has been
implemented by using nuclear magnetic resonance (NMR) techniques for
a system with four states\cite{Jones} and more recently using
quantum optical methods\cite{Scully}. An efficient quantum algorithm
for global optimization based on such Grover's search procedure can
find applications in a wide range of fields\cite{alan2}.

\begin{figure}
\begin{center}
\includegraphics[width=0.6\textwidth,height=0.75\textheight]{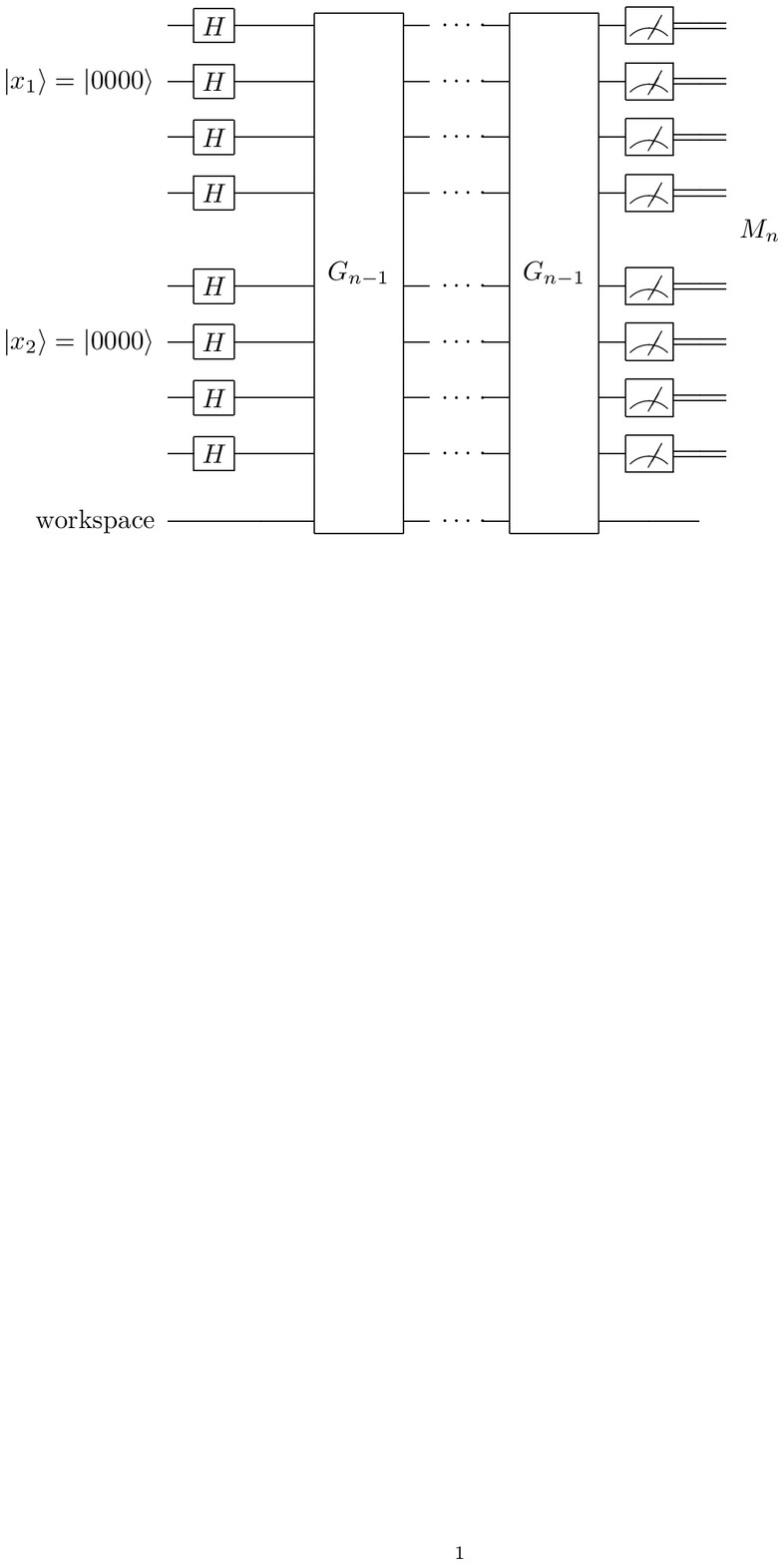}
\end{center}
\caption{Quantum circuit for searching the global minimum. The
$G_{n-1}$ is the oracle to rotate the vector toward the state with
$f(x_1,x_2,...) \le M_{n-1}$, where $M_{n-1}$ is the minimum of all
measurement results. The registers will be initialize to
$|x_1\rangle=|0000\rangle$ and $|x_2\rangle=|0000\rangle$ for
variable $x_1$ and $x_2$. After using the Hadamard gates to convert
the initial state into the superposition state, the adapted Grover
operators will be applied to rotate the superposition state to the
specific state. The threshold value of Grover operator will be
updated based on the measurement result after certain number of
rotations. \label{circuit}}
\end{figure}

In this paper, we will demonstrate that a modified Grover's quantum
algorithm can be applied to real problems of finding a global
minimum using modest numbers of quantum bits. We will simulate the
revised quantum search algorithm using a  classical computer. The
limitation of computer resource such as the memory and speed of CPU
will prohibit a large scale quantum computer algorithm  simulation
on a classical computer. Thus, first we implement the algorithm for
simple test functions and small size Lennard-Jones clusters. The
quantum search circuit is shown in Fig. \ref{circuit}. In this
figure, we divided the register into three groups, where the
Hadamard gates are operated on the  initialized  registers, then we
applied  the Grover iteration to rotate the superposition states
into target states. The measurement result after the iteration is
used to update the threshold value in the Grover iteration steps.
Grover oracle is replaced by the adapted threshold to search all the
states which $f(x_1,x_2,..)\le M_{n-1}$, where $M_{n-1}$ gives the
minimum measurement value. The subscript $n$ means the steps of
measurements and $M_{n}$ means the value of the $n^{th}$
measurement. The iteration number during each search will be
preselected. After each iteration, the result will be measured and
compared with the other measured results to setup a new threshold
value.

\begin{figure}
\begin{center}
\includegraphics[width=1.0\textwidth]{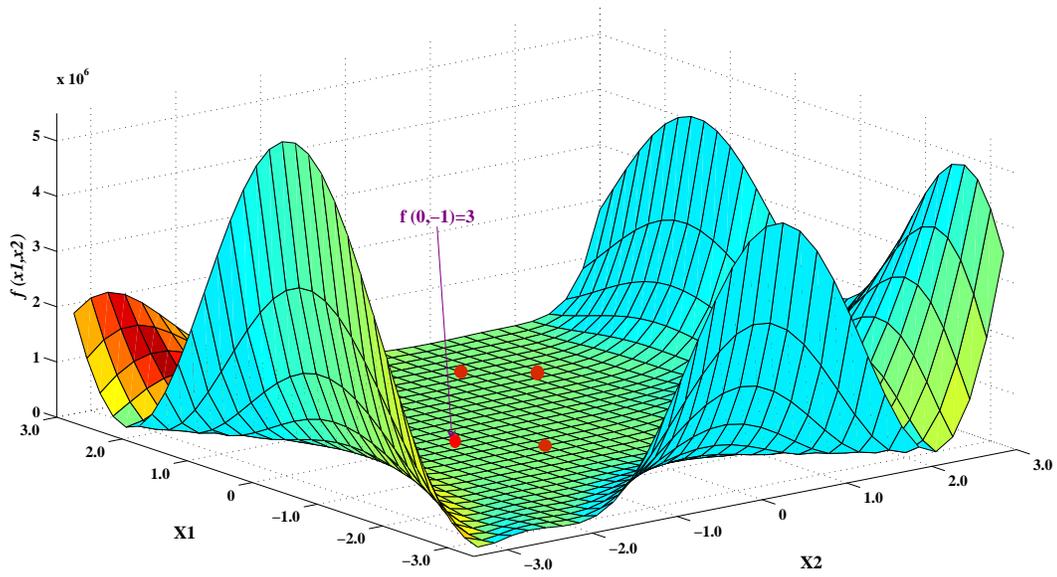}
\end{center}
\caption{The potential surface near the global minimum  of the GP
function (See the text). It is difficult to tell the four minima
points visually on a scale of zero to five million, so I marked them
by brown circles. $f(0,-1)=3$ is the global minimum which was
indicated by the arrow. \label{classicGP}}
\end{figure}

\begin{figure}
\begin{center}
\includegraphics[height=0.5\textheight]{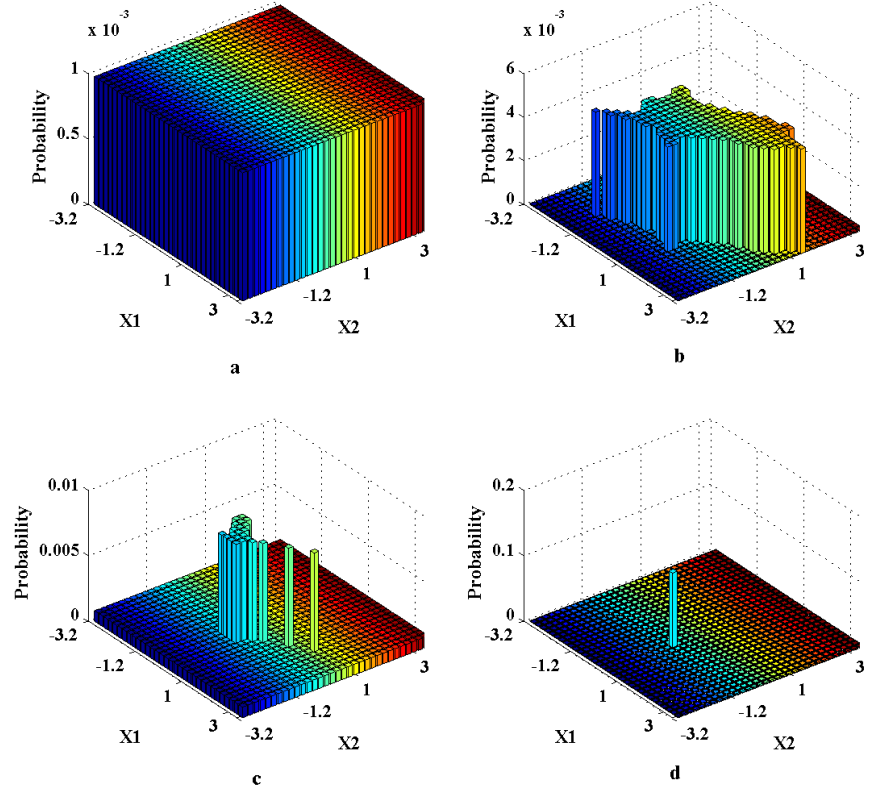}
\end{center}
\caption{Probability distribution of the state function before the
measurement for the GP function. The panel a is the initial state
corresponding to the superposition of all possible states, the
panels b, c and d are the distributions of step 5, 10 and 16
respectively.
\label{gpFig}}
\end{figure}

We first applied  the adapted quantum search algorithm to test
a simple  analytical function used in global optimization: the
Goldstein-Price function(GP) which is given by\cite{Kais2-Opt}
\bea
f(x_1,x_2)&=&[1+(x_1+x_2+1)^2(19-14x_1+3x_1^2-14x_2+6x_1x_2+3x_2^2)]
[30+(2x_1-3x_2)^2\nonumber\\
& &(18-32x_1+12x_1^2+48x_2-36x_1x_2+27x_2^2)], \eea where $-2\le x_i
\le 2$. The GP function is an excellent test function for any global
optimization method, which has four local minima in the whole
region. One global minima is located at $(0, -1)$ with the function
equals to 3, the other three local minima are $f(-0.6,-0.4)=30$,
$f(1.8,0.2)=84$ and $f(1.2,0.8)=840$. The potential surface near the
global minimum is shown in Fig. \ref{classicGP}. It is a kind of
difficult to visual minima on a scale of $0$ to five million. We
added four brown circles to mark all local minima positions,
furthermore, the global minima $f(x_1,x_2)=3$ is shown by arrow in
Fig. \ref{classicGP}. We used 10 qubits as the registers (As an example,
see the Appendix for detailed calculations using 2 qubits).
The registers will be divided into two groups
 to present the variable $x_1$ and $x_2$. The searching range is
 $x_{1,2}\in [-3.2,3.0]$, which covers all local and global minima.
 After applying the Hadamard gate, the registers group will be
initialized into the superposition state, each will be used to cover
$2^5$ discrete points in the searching range, namely, each basis
function will be mapped to  the number within the searching domain.
Then, the measurement will be performed to obtain the first
threshold value after selected number of Grover iterations was
applied. The number of iterations before each measurement is
important to reduce the total iteration number. We chose the
sequence: $0, 0, 0, 1, 1, 0, 1, 1, 2, 1, 2, 3, 1, 4, 5, 1, 6, 2, 7,
9, 11, 13, 16, 5...$, as the iteration number before each
measurement, i.e. for step 1 we measure the state without any Grover
iteration, for step 4, we measure the state after one Grover
iteration. This sequence has been proposed in Ref. \cite{BARITOMPA}
to reduce the Grover iteration numbers for adopted Grover search
method. During each iteration, the state function will be rotated
towards the threshold value, which is always the best measurement
result at previous steps. The new obtained value will be compared
with the old threshold value. The threshold value will be reset to
the new value if the old threshold value is larger than the new
measurement result, otherwise it is unchanged. The iteration will
continue until convergence  is reached. The quantum search yields
the same result(high probability) as the classic method with
16 steps. In Fig. \ref{gpFig}, we show the probability distribution
of the state function before the iteration steps. The top left panel
is the initial state, where the state function is the superposition
for every possible state, and the measurement result yields
379605.8306 after step one without any Grover iteration. In
Fig.\ref{gpFig},  panel (b) is the state function after total 2
Grover iterations at step 5, where the measurement result yields
4038.3764. As we can see the probabilities for smaller function
values become larger, meanwhile the eigenfunction corresponding to
larger function values starts decreasing. In panel (c), the area of
high probability reduced a lot compared with panel (b). It means
that the search keeps converging at step 10. At step 16 with total
22 Grover iteration, we reach the global minimum value 3 at $x_1=0$
and $x_2=-1$ as shown in panel (d) of Fig. \ref{gpFig}.

\begin{figure}
\begin{center}
\includegraphics[width=1.0\textwidth,height=0.5\textheight]{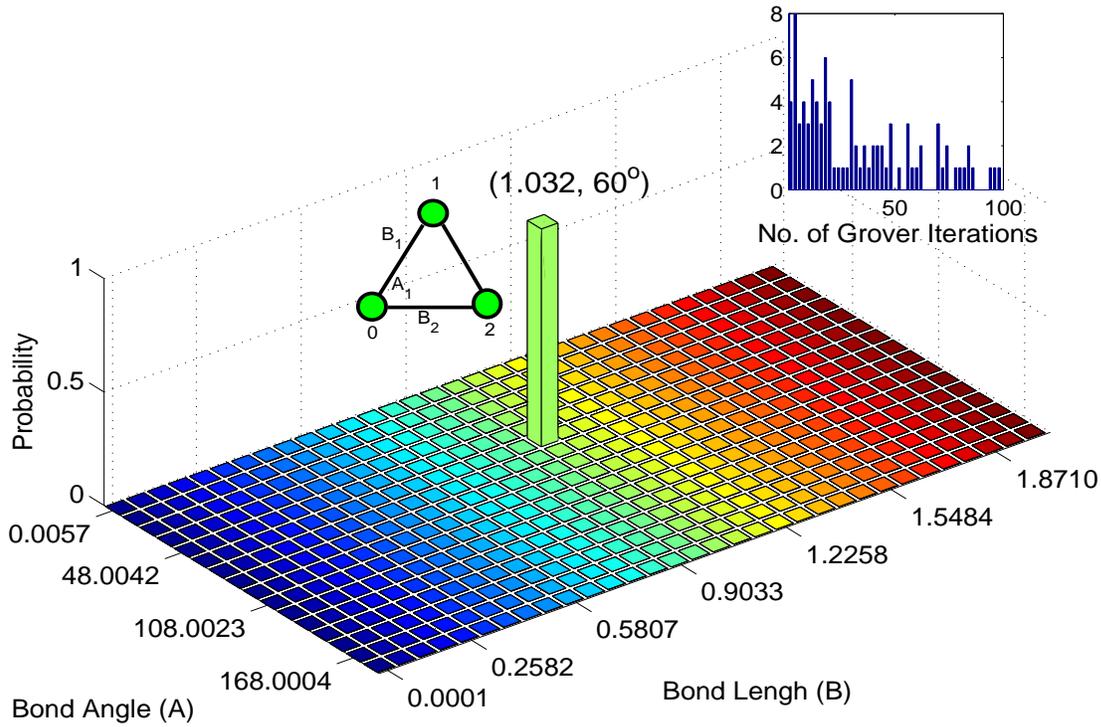}
\end{center}
\caption{Final probability distribution of the state function for LJ
(N=3). The global minimum located at $B_1=1.0323,B_2=1.0323$ and
$A_1=1.0473 rad=60^o$. The top panel is the distribution of total
measure step before reaching the global minimum for 100 search
results. The global minimum and corresponding structure are also
shown in figure. \label{LJ3DFig}}
\end{figure}

Let us further  illustrate this approach by considering a real and practical optimization problem:
 finding the global minimum of Lennard-Jones clusters, clusters of atoms or molecules
 that interact with each other through the Lennard-Jones potential. The Lennard-Jones potential
(referred to as the L-J potential or 6-12 potential) is a simple
mathematical model that describes the long range attractive van der Waals force
and the short range Pauli repulsion force.
 The L-J potential is of the form:
\beq
 V_{LJ}(r) \,=\, \epsilon [ (\frac{2^{1/6}\sigma}{r})^{12} \,-\, (\frac{2^{1/6}\sigma}{r})^{6}],
\eeq where $\epsilon$ and $2^{1/6}\sigma$ are the pair equilibrium
well depth and separation, respectively. $r$ is the relative distance
between two particles. We will employ reduced units in our simulation
and define $\epsilon\,=\,2^{1/6}\sigma\,=\,1$.
 The L-J potential is a relatively good approximation and due to
 its simplicity often used to describe the properties of gases, and
 to model dispersion and overlap interactions in molecular models.
 It is particularly accurate for noble gas atoms and is a good
 approximation at long and short distances
 for neutral atoms, molecules and clusters.
Lennard-Jones clusters are excellent for testing the efficiency of
global optimization algorithms \cite{ljcluster}.  Homogeneous
Lennard-Jones clusters have well-established minima and regular
minimum-energy structures for very large clusters \cite{LJ-10a}.
However, the number of local minima apparently grows rapidly and
finding the global minimum in Lennard-Jones clusters is an NP-hard
problem \cite{LJ-12}. Several global optimization methods have been
applied to the energy function of Lennard-Jones clusters. The total
energy for a Lennard-Jones cluster of M particles is: $E_M \,=\,
\sum_{i=1}^{M-1} \; \sum_{j=i+1}^{M} V_{LJ}(r_{ij})$, where $r_{ij}$
is the distance between the $i$-th and the $j$-th particles and
$V_{LJ}(r)$ is the Lennard-Jones two-body potential. We start
simulating the process of searching the global minimum for $M=3$
particles. A total of 9 register qubits (N=$2^9$=512 mesh points)
were separated into two groups for presenting three variable $B_1,
B_2$, and $A_1$, where $B_1$ is the bond length between the atom 0
and 1, $B_2$ is the bond length between atom 0 and 2, and $A_1$ is
the bond angle of atom 0, 1 and 0, 2 as shown in Fig. \ref{LJ3DFig}.
Five qubits will be used as the first register to cover the space
$B_1=B_2$, and four qubits will be used as the second register for
$A_1$. The searching range for $B_{1,2} \in [0.0001,2]$ and for $A_1
\in [0.0001,\pi]$. The quantum search yields the minimum value
-2.9094 at $B_1=1.0323$, $B_2=1.0323$ and $A_1=1.0473 rad=60^o$.
Compared with the classical minimum potential , it is slightly
higher. This is due to our mesh can not cover the exact minimum
value. Unlike the search method for GP function, where the number of
iterations is preselected based upon the proposed sequence, here we
increase the number of iterations after every measurement to study
the importance of the iteration sequence. The number of iterations
increases as $1,2,3,4...$ In Fig. \ref{LJ3DFig}, we show the search
results and the total iteration steps. From the histogram of total
number of iterations
 for 100 independent searches, we found that the average  number of iteration
is about 21. This indicates that the running time of adapted search
algorithm (a total of 231 iterations) is still the same as the
Grover search algorithm $O(\sqrt{N})$, which is about $10\sqrt{2^9}$
in this example. The configuration corresponding to the minimum for
LJ cluster is also shown in  Fig. \ref{LJ3DFig}.

\begin{figure}
\begin{center}
\includegraphics[width=0.9\textwidth,height=0.8\textheight]{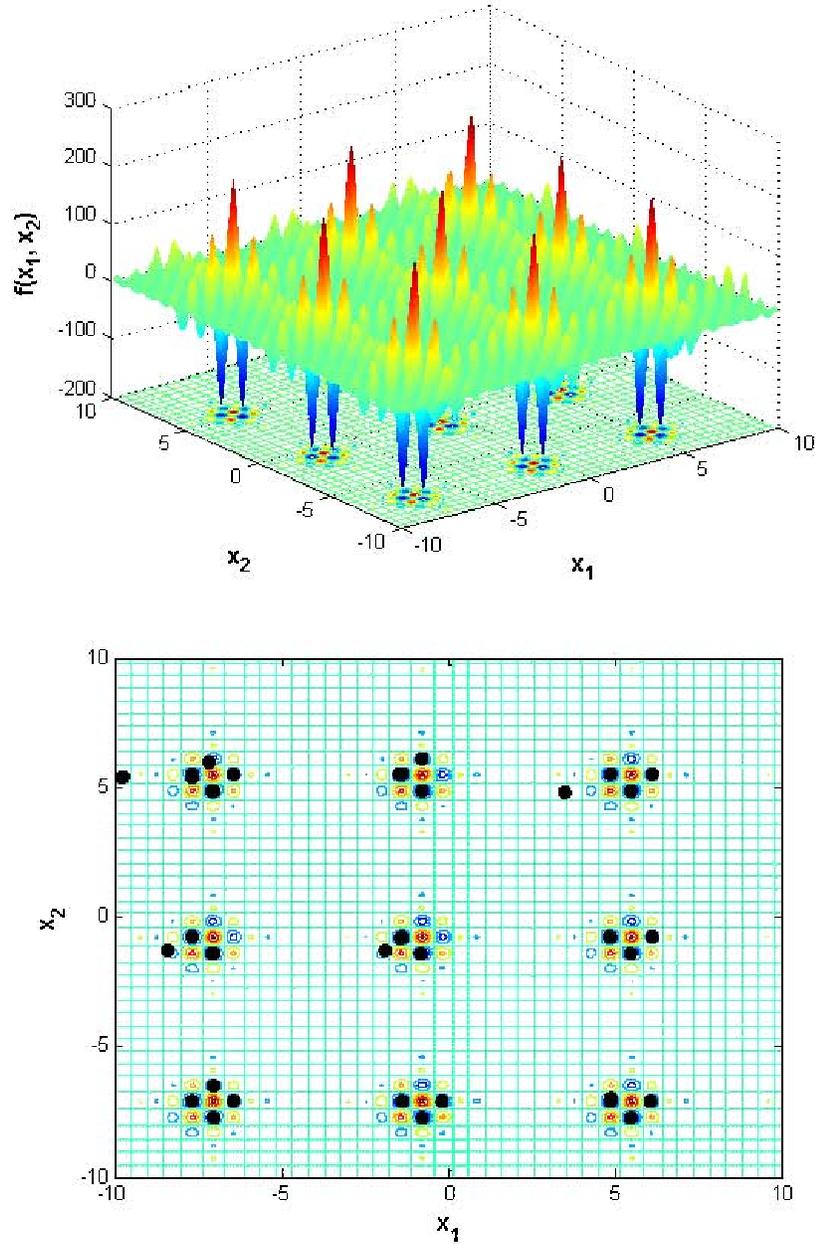}
\end{center}
\caption{The surface potential of the Shubert function and the final quantum
search results. The top panel is the surface potential for the
Shubert function with the range $x_{1,2}\in(-10,10)$. The bottom
panel is the contour of the function with the quantum search
results. The black dots present the measurement results for all
search steps.
\label{SurShub}}
\end{figure}

In order to expand the adapted quantum search algorithm to search
the global minimum for larger number of variables and to overcome
the limit of using large number of qubits in the computation, we
 combined the classical  pivot search method\cite{Kais1-Opt,Kais2-Opt}
with the quantum Grover's  search algorithm. The basic scheme is as
following: {\it Step (1)}: Generate $N$ random probes, where $N$
equals to $2^{number\; of \; qubits}$, then shift it into
superposition of the entire state space.
 \emph{Step (2)}: Use the quantum Grover algorithm
mentioned before to do the comparison. Select and keep about the
smallest 15\% of the original $N$ random probes as pivot probes.
\emph{Step (3)}: Initialize the quantum computer with the state
associated with these pivot probes, apply a series of controlled
Hadamard gates to produce the superposition state with points near
the selected probes. $ x_{R,i}=x_{B,i}+\Delta x_i$, where $\Delta
x_i$ is a randomly generated vector according to a particular
distribution such as Gaussian distribution \cite{Kais1-Opt}.
\emph{Step (4)}: Redo \emph{Step (2)}and keep going until the
criteria of convergence is satisfied.
 Using this procedure, it is possible
to cover the entire searching space by a small number of qubits.
Moreover, this small number of qubits is sufficient to cover each
subdomain to yield the desired resolution. To illustrate this
combined approach, we search the global minimum for the Shubert test
function, which is given by\cite{Kais2-Opt}:
 \beq
f(x_1,x_2)=[\sum_{i=1}^{i=5}i
\cos[(i+1)x_1+i]]\times[\sum_{i=1}^{i=5}i \cos[(i+1)x_2+i]] \eeq
with $-10\le x_{1,2}\le 10$, which has 760 local minima, 18 of which
are global with $f(x_1,x_2)=-186.7309$. The surface potential of
this function is shown in
 Fig.\ref{SurShub}. Ten qubits were used to do this simulation, each $x$
 was assigned 5 qubits. Following the same procedure mentioned above,
 we initially generated $2^{10}$ random points.
 Then 15\% minimum of these points were picked up by quantum Grover algorithm  as pivots.
 After that, we arranged the pivots based on the weight of the optimized function
 ($\exp(-f(x_1,x_2)/kT)$), where $kT$ is just a fixed parameter and equals to 50 in our simulation.
 We generated the other points according to the
 Gaussian distribution. We ran this simulation for 98 times, the
 researched minimum values are between -30.56 and -186.73. Over 80\% points
 are located at -186.73, which is the exact global minimum for this
 function. It also covers all 18 global minima and the average iteration is 1300.
  The simulation results are  shown in Fig. \ref{SurShub}, where the black dots are
measurement results on the contour of the surface potential with red
dots as global minima. It seems this method only converges to the
global minima 80\% of times.
However, the advantage is that  only 10 qubits used, which means $2^5$ pivot
points for each x axis,  with total $2^{10}$ mesh points.
It is a very small mesh based on current computer. If we can use 20
qubits (10 for each axis), the number of mesh points will be much
larger. Under this situation, the convergence speed will be much
faster.

\begin{figure}
\begin{center}
\includegraphics[width=1.0\textwidth]{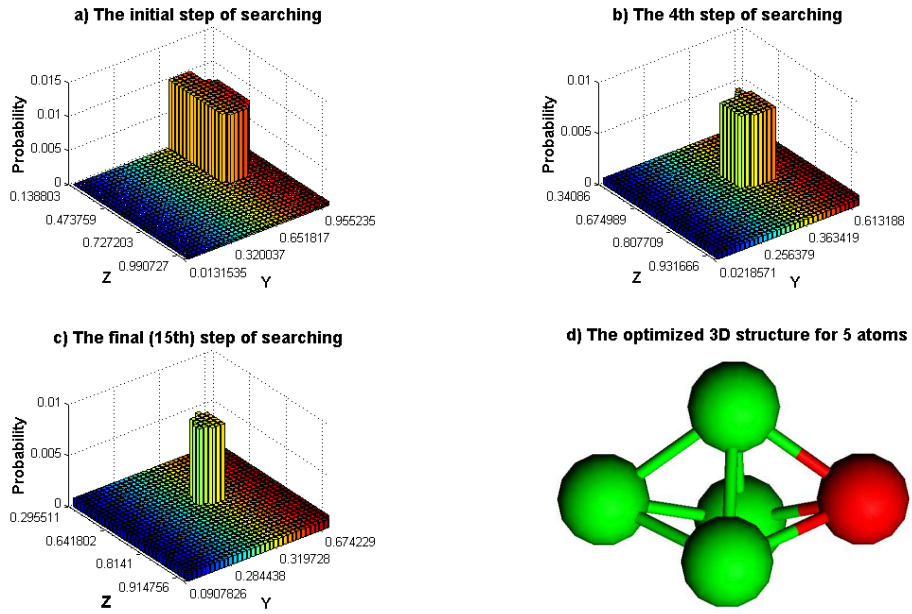}
\end{center}
\caption{Probability distribution of the state function for LJ
cluster (N=4). The a, b and c are the steps during the searching.
The global minimum energy is -5.9926 and the coordinates for the 4th
atom are (0.0, 0.28444, 0.81344). Although there are many bars, they
are quite close to each other. d is the optimized structure for 5
atoms cluster with the energy of -9.0952. The green atoms are fixed
atoms, while the red one is the free atom. The distances between the
red and the green ones are 0.99, 0.99 and 1.00 separately.
 \label{structure}}
\end{figure}

Furthermore, following the same steps, we also applied this combined
method to search optimized structure for LJ cluster. We tried 5
atoms and got the exact same results as the classical method.
 The detailed procedure for the 3 atoms simulation is as following:
 we first set 5 qubits for $B_{1,2}$ and 5 qubits for $A_1$. The same
 previous range, which is $B_{1,2} \in [0.0001,2]$ and for
 $A_1\in[0.0001,\pi]$. We first generated $2^5$ random $B_{1,2}$ and $2^5$ random
 $A_1$ in the above range. Then we used the combined classical pivot
 method and Grover's search algorithm which was mentioned in
 previous paragraph. After the search, we got the global minimum
 structure for the 3 atoms cluster (the same structure shown in Fig. \ref{LJ3DFig}).
 The distance between each atom is 0.99889 and the total potential for this structure is
 -2.9999, which is almost the same as the classical result(-3.0).
After that, based on the optimized structure of 3 atoms, we added another atom
 to form the 4 atoms cluster. We fixed the original 3 atoms and set the forth one
 free. We used $X, Y$ and $Z$ to express the coordinates for the forth
 atom and used 10 qubits in the simulation,  the
 same number as in the previous simulations. We tried two different ways to perform the
 simulations: In the first method, we let the forth atom totally free in $X, Y$ and $Z$
 directions and set 4 qubits for $X$
 axis, 3 qubits for $Y$ axis and the rest 3 qubits for $Z$ axis. In
 order to save simulation steps, the ranges for all coordinates are
 $X \in [-0.5, 0.5]$, $Y \in [0.01, 1.01]$ and $Z \in [0.01, 1.01]$.
 Then we followed the same steps as 3 atoms, generated random points
 and started the search. After a number of simulations, the result converged to
 -5.0 in average, which is not quite accurate (classical=-6.0). The second method is
 fixing $X$ at 0.0 and arranging 5 qubits for $Y$ and $Z$ respectively.
 The ranges for $Y$ and $Z$ are the same,  between 0.01 and
 1.01. The rest of the  procedures is  the same as for 3 atoms. the simulation
results gives
  the lowest potential for this cluster is -5.9926, which is quite close
 to the classical result (-6.0). The probability distribution for the
 whole procedures is shown in Fig. \ref{structure}, as well as the final structure.
 We stepped forward to use this
 method to simulation the most optimized structure for 5 atoms
 cluster. The simulation procedure is the same as for 4 atoms
 cluster.  We fixed the previous optimized structure and added the
$5^{th}$ free atom. We also used the above two methods,  results
show that the second method gave more accurate results. We obtained the value
 -9.0952, which is  close to the classical simulation value
(-9.103852).  Although we did not  get the exact
optimized structures  with the current simulations but with more qubits
one should cover the exact results.

 It is known that
the $\sqrt{N}$ is the optimal running time for quantum search
algorithm. The combined search method does not reduce the total
rotation steps, but does reduce the required number of qubits to do
the simulation.
 Due to the limited available qubits in the classical computer,
  we can only set one atom free with all other atoms fixed.
 However, in a quantum computer,
  with enough qubits available, we can perform full optimization
for all atoms. For example,
 for larger LJ clusters, if we had larger qubits, we
can incorporate the partial knowledge that we had by starting with
the structure of the smaller ($M-k$) clusters and adding  $k$
additional particles at random\cite{Kais1-Opt,Northby}. In a
previous work\cite{Kais1-Opt}, using the pivot method we have shown
that the computational cost (CPU time) scales as $M^{2.9}$ with the
number of L-J particles to be minimized. With various practical
improvements, if one reduces the scale to $~M^2$. If we can assign a
certain number of qubits for each particle, then the Grover
algorithm will reduce the search steps into the order of $O(M)$ for
this specific case. In any `growing' problem, such as minimum energy
configuration of clusters, self-avoiding walks, protein folding,
etc., this systematic approach to solving the structure of large
clusters can be incorporated. One of the powerful features of this
combined algorithm is that information such as this can be built
into the initialization of the probes.

We have used an adapted quantum search algorithm to search the
global minima for test functions and LJ clusters. Our quantum
computer simulations on the classical computer yield the same global
minimum values as the classical search method with high probability.
We also show how to combine the classical Pivot method with the
adapted quantum search algorithm to search for the global minimum in
larger domains.
 Recently, Jordan\cite{jordan} proposed a fast quantum algorithm for estimating numerical
 gradients
with one query. One can use this method to search the potential gradient with zero
value. This will rotate the entire space towards the state function which corresponds
to all minima. The measurement will yield one of the minima in
stead of any point in the search domain. Combining this  with our search
method will greatly reduce the number of rotations needed for finding the  global minimum.
With further improvements in the quantum search algorithms, we expect to see
solutions of previously intractable global optimization problems
in many different fields.\\

In summary, the manuscript contains novel results and a
proof-of-principle about the use of quantum computers for the
simulation of a global minimum. First, the paper provides
fundamental insight   into the quantum simulation of global
optimization problems, and second we implemented some simple
applications. We also demonstrated for the first time  that a
modified Grover's quantum algorithm can be applied to real problems
of finding a global minimum using a modest number of qubits. If a
quantum computer that would allow for these calculations to be
carried out were available now, we believe that the development of
algorithms for optimization is of great importance in many practical
fields and further motivates the construction of these devices.

Experimentally, the Grover's algorithm has been demonstrated by
nuclear magnetic resonance (NMR) \cite{nmr1, nmr2, nmr3,spec,expm}
and quantum optics \cite{Scully} for small number of qubits.
Although it is easy to obtain the Grover's oracle by classic
computers, it is very hard to realize this oracle in quantum
circuit. There are no efficient universal methods to design this
oracle until now. Recently there are a few attempts to solve this
problem directly. Ju and coworkers \cite{Ju} implemented Grover's
oracle function by Boolean logic in quantum circuit. However, they
used N ($2^{number\; of \; qubits}$) Boolean logic to represent the
oracle, which makes the circuit design not efficient. On the other
hand, Xu and coworkers \cite{Xu} successfully used the adiabatic
search algorithm to realize Grover's algorithm without oracle by
encoding the database to quantum format and forming the problem
Hamiltonian form target value. Further research is still needed to
overcome the Grover's oracle.

\begin{acknowledgements}

We thank Jonathan Baugh for useful discussions and the Army Research
Office (ARO) for funding.

\end{acknowledgements}

\newpage
  \begin{center}
    {\bf APPENDIX}
  \end{center}

This Appendix show how the modified Grover algorithm is used for the
simulation of the global minimum of the GP function. In the text, we
used a total of 10 qubits for the optimization. Here as an example
we provide results for only two qubits. One qubit is used to
represent the $x_1$ axis and the other for the $x_2$ axis. The
search range was $x_{1,2}\in [-3.2,3.0]$ and discretized into
$2^{1}=2$ points on each axis.

{\it Step (1)}: Perform the Walsh-Hadamard transformation to place
the system into a superposition with equal probabilities for all
states. The obtained vector is also our original source $\sour$. The
formula for this step is
  $\sour \,=\,
  \overset{2}{\underset{i=1}{\bigotimes}}
 H \ketz$. Thus, we obtain the  vector $\sour = (0.5\, \,0.5\, \,0.5\, \,0.5)^{T}$.

{\it Step (2)}: Generate the $P_s$ operator, $P_{s} = 2\sour\asour
\,- I\,$, where $\sour$ is the vector we obtain in Step 1. $P_s$
increases the amplitude of the selected state and takes the form
\beq P_s = \frac{1}{2} \left(
\begin{array}{cccc}
-1 & 1 & 1 & 1\\
1 & -1 & 1 & 1\\
1 & 1 & -1 & 1\\
1 & 1 & 1 & -1\\
\end{array}
\right) \eeq

{\it Step (3)}: Obtain the entry matching the search criterion
$\targe$ (can also be called the target source). There is no direct
way to  obtain the target source from the pure quantum method, so a
"black box" type is used in the current simulation. Thus,  it was
obtained by an indirect mapping. We calculated values of the GP
function at all mesh points and picked the lowest one by classical
comparison. Then we marked the corresponding part of $x_{1,2}$ as 1
and the rest  as zeros. Followed by applying the Kronecker tensor
product to obtain the target source $\targe$, which is a similar
vector to $\sour$. For this example, the point $(-3.2, -3.2)$ has
the lowest value. The corresponding target source is $\targe = (1\,
\,0)^{T}{\bigotimes}(1\, \,0)^{T} = (1\, \,0\, \,0\, \,0)^{T}$.

{\it Step (4)}: Generate the $P_t$ operator, where $P_t =
 \,I - \, 2\targe\atarge$. This operator reverses the selected
 state and takes the form
 \beq P_t =
\left(
\begin{array}{cccc}
-1 & 0 & 0 & 0\\
0 & 1 & 0 & 0\\
0 & 0 & 1 & 0\\
0 & 0 & 0 & 1\\
\end{array}
\right) \eeq

{\it Step (5)}: Apply the Grover operator, $G=P_{s}P_{t}$, $O(\sqrt{N})$
times to the superposition state ($\sour$). The new $G$ operator has
the effect of both $P_s$ and $P_t$, which reverses the selected
state first and then increases its amplitude. For this example, it takes the form
 \beq G\sour = P_{s}P_{t}\sour = P_{s}
\left(
\begin{array}{c}
-0.5\\ 0.5\\ 0.5\\ 0.5\\
\end{array}\right)
= \left(
\begin{array}{c}
1\\ 0\\ 0\\ 0\\
\end{array}\right) \eeq

For this simple example, we obtained the correct answer in one step.

\newpage


\end{document}